\documentclass[%
 reprint,
 amsmath,amssymb,
aps,
]{revtex4-1}

\usepackage{graphicx}
\usepackage{dcolumn}
\usepackage{bm}
\usepackage{epstopdf}

\begin{document}

\preprint{APS/123-QED}

\title{Dispersion regions overlapping for bulk and surface polaritons in a magnetic-semiconductor superlattice}

\author{Volodymyr I. Fesenko}
\email{v.i.fesenko@ieee.org}
\author{Vladimir R. Tuz}
\email{tvr@rian.kharkov.ua}
\affiliation{
 Institute of Radio Astronomy of National Academy of
Sciences of Ukraine, \\
4, Mystetstv St., Kharkiv 61002, Ukraine
}
\author{Illia V. Fedorin}
\email{fedorin.ilya@gmail.com} \affiliation{
 National Technical University `Kharkiv Polytechnical
Institute',  \\
21, Polytechnichna St., Kharkiv 61002, Ukraine, }

\date{\today}

\begin{abstract}
Extraordinary dispersion features of both bulk and surface
polaritons in a finely-stratified magnetic-semiconductor structure
which is under an action of an external static magnetic field in the
Voigt geometry are discussed in this letter. It is shown that the
conditions for total overlapping dispersion regions of simultaneous
existence of bulk and surface polaritons can be reached providing a
conscious choice of the constitutive parameters and material
fractions for both magnetic and semiconductor subsystems.
\end{abstract}


\maketitle


The polariton is introduced as a quasi-particle which characterizes
a coupling between electromagnetic waves (photons) and a diverse
variety of dipole-active elementary excitations inherent to a matter
such as phonons, plasmons, excitons, etc. Although the concept of
quasi-particles is related to quantum mechanics, the polariton can
be considered  as a \textit{macroscopic} phenomena concerning on
interaction of electromagnetic waves with macroscopic normal modes
(eigenwaves) of a matter assuming their wavelength are long enough
so that the medium can be treated as a continuous one. In such a way
the theory of polaritons is developed without specifying which kind
of dipole excitation is coupled to electromagnetic waves, because
the specific nature of this excitation is completely defined only by
the dielectric function (e.g. permittivity) of a medium
\cite{Mills_RepProgPhys_1974}. This approach implies a particular
consideration of polaritons in a bulk material (bulk polariton) and
on its surface (surface polariton).

Although the nature of these two particular modes is the same and is
related to the medium polarizability, surface polaritons are
distinguished from bulk polaritons by the fact that their amplitudes
decay exponentially away from the surface in the direction normal to
it \cite{ushioda_1981}. This means that the normal component of the
wavevector of  a surface polariton is purely imaginary and
consequently it cannot propagate away from surface. Therefore, these
surface modes do not couple linearly with bulk polaritons either
inside or outside the surface. In particular, this fact manifests
itself in the different areas of existence of bulk and surface
polaritons on their dispersion characteristics.

As already mentioned, from the macroscopic viewpoint, the dispersion
conditions for bulk and surface polaritons are defined by the
dielectric function of a medium, and in the case of surface modes
the difference in signs of the dielectric functions of patterning
materials is required (i.e. the real parts of the permittivity
scalars of two patterning materials must have opposite signs).
Nevertheless, if at least one of two patterning materials is
anisotropic  (due to its crystalline nature or as a result of an
external action, such as application of the static magnetic field)
the permittivity appears as a tensor quantity, and the dispersion
characteristics of polaritons become to be more complicated
\cite{Burstein_PhysRevLett_1972, Burstein_PhysRevB_1974}, e.g. the
propagation of surface polaritons can appear to be permissible, even
though the real parts of all components of the permittivity tensors
of both materials are positive \cite{Crasovan_OptLett_2005}.

Besides the dielectric function, the properties of polaritons can
also be determined by the magnetic function in a magneto-optically
active medium in which electromagnetic waves can be coupled to spin
waves in (anti)ferromagnetic \cite{Kaganov_PhysUsp_1997}. Apart from
flexible modulation of the polaritons  properties by an external
magnetic field, surface magnetic polaritons also promise
nonreciprocal effect and multi-bands of propagation
\cite{Burstein_JPhysC_1973}. Moreover, in such systems it was
demonstrated that the bulk and surface polaritons can exist in the
same frequency range, but with different wavevectors
\cite{Camley_PhysRevB_1982, Tuz_MAGMA_2016}, as well as the
possibility of obtaining very narrow frequency range where a surface
polariton branch can merge into the continuum of bulk modes is found
out \cite{Tagiyeva_JPCS_2007, Tagiyeva_J_Supercond_Nov_Magn_2012}.

It is obvious that the simultaneous combining together of dielectric
and magnetic functions into a single electromagnetic (or generally,
gyroelectromagnetic) system can bring a number of unique dispersion
features of polaritons that are unattainable in separate subsystems.
In particular, in this letter our goal is to demonstrate, for the
first time to our knowledge, that the dispersion regions of bulk and
surface polaritons can overlap in a composite magnetic-semiconductor
structure providing a proper selection of the constitutive
parameters and material fractions for both magnetic and
semiconductor subsystems.

\begin{figure}
\centering \fbox{\includegraphics[width=6.3cm]{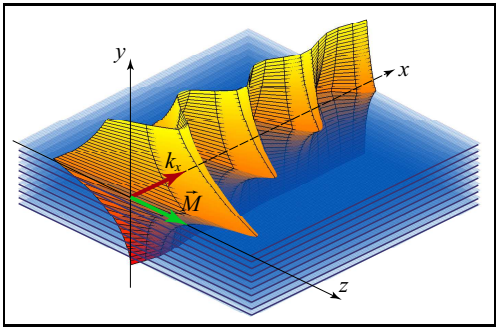}}
\caption{(Color online) The problem geometry and a visual
representation of the tangential field distribution ($E_z$) of the
surface polariton propagating on a surface of a
magnetic-semiconductor superlattice which is under an action of an
external static magnetic field applied in the Voigt configuration.}
\label{fig:fig_SPP}
\end{figure}

Therefore, in this paper we study dispersion characteristics of both
bulk and surface polaritons propagating in a \textit{semi-infinite}
stack of identical composite double-layered slabs arranged along the
$y$-axis (Fig.~\ref{fig:fig_SPP}). Each composite slab within the
stack includes magnetic (with constitutive parameters
$\varepsilon_m$, $\hat \mu_m$) and semiconductor (with constitutive
parameters $\hat \varepsilon_s$, $\mu_s$) layers with thicknesses
$d_m$ and $d_s$, respectively. Thus, the stack possesses a periodic
structure (with period $L = d_m + d_s$) that fills half-space $y<0$
and adjoins a vacuum occupying half-space $y>0$. We suppose that the
structure is a \textit{finely-stratified} one (i.e. it is a
superlattice), whose characteristic dimensions $d_m$, $d_s$ and $L$
are all much smaller than the wavelength in the corresponding layer
$d_m\ll \lambda$, $d_s \ll \lambda$, and period $L \ll \lambda$ (the
long-wavelength limit). Along the $x$ and $z$ directions the system
is considered to be infinite.  An external static magnetic field
$\vec M$ is directed along the $z$-axis provided that the strength
of this field is high enough to form a homogeneous saturated state
of magnetic as well as semiconductor subsystems.

Taking into account the smallness of layers' and period's
thicknesses compared to the wavelength, the standard homogenization
procedure \cite{Agranovich_SolidStateCommun_1991,
Eliseeva_TechPhys_2008, Tuz_MAGMA_2016} from the effective medium
theory is applied in order to derive averaged expressions for
effective constitutive parameters of the finely-stratified structure
under investigation. From the viewpoint of this theory, the
composite multilayered structure is approximately represented as an
\textit{anisotropic uniform} medium in which one optical axes is
directed along the structure periodicity (the $y$-axis) while the
second one coincides with the direction of the external magnetic
field (the $z$-axis). Therefore, the resulting composite medium is
characterized with two tensors of relative effective permittivity
$\hat\varepsilon_{eff}$ and relative effective permeability
$\hat\mu_{eff}$ obtained in the form:
\begin{equation}
 \hat \varepsilon_{eff}=\left( {\begin{matrix}
   {\varepsilon_{xx}} & {\varepsilon_{xy}} & 0 \cr
   {-\varepsilon_{xy} } & {\varepsilon_{yy} } & 0 \cr
   0 & 0 & {\varepsilon_{zz}} \cr
\end{matrix}
} \right),~~~~~ \hat\mu_{eff}=\left( {\begin{matrix}
   {\mu_{xx} } & {\mu_{xy}} & 0  \cr
   {-\mu_{xy} } & {\mu_{yy}} & 0  \cr
   0 & 0 & {\mu_{zz} }  \cr
 \end{matrix}
} \right), \label{eq:eff}
\end{equation}
where expressions for the tensors components derived via underlying
constitutive parameters of magnetic ($\varepsilon_m$, $\hat \mu_m$)
and semiconductor ($\hat \varepsilon_s$, $\mu_s$) layers as well as
their dispersion characteristics one can find in
\cite{Tuz_MAGMA_2016, Tuz_JOpt_2010, Tuz_PIERB_2012, Tuz_JO_2015,
Tuz_Springer_2016}. Remarkably, all diagonal elements of both
tensors appear to be different ($\mu_{xx}\ne \mu_{yy}\ne \mu_{zz}$,
$\varepsilon_{xx}\ne \varepsilon_{yy}\ne \varepsilon_{zz}$), so the
structure under investigation exhibits properties of a
\textit{biaxial bigyrotropic} (gyroelectromagnetic) crystal.

In accordance with the geometry of the problem (see,
Fig.~\ref{fig:fig_SPP}), the static magnetic field $\vec M$ is
directed along the $z$-axis, so the system is considered to be in
the Voigt configuration. In such configuration the magnetic field
vector in the TM mode (that has field components $\{E_x,E_y,H_z\}$)
is parallel to the external magnetic field $\vec M$, which results
in the absence of its interaction with the magnetic subsystem
\cite{Tarkhanyan_OptExpress_2006, Tarkhanyan_JMMM_2010}. Thus,
further dispersion features only of the TE mode (that has field
components $\{H_x,H_y,E_z\}$) are of interest.

Involving a pair of the curl Maxwell's equations for time-harmonic
fields (a time factor is considered to be in the form $\exp(-i\omega
t)$) for the wave propagating along the $x$-axis  in a standard way
\cite{Burstein_JPhysC_1973, Tuz_MAGMA_2016} one can obtain the
following relation between frequency and wavenumber $k_x$:
\begin{equation}
k_x^2 = k_0^2\varepsilon_{zz}\mu_{yy}\mu_v/\mu_{xx},
\label{eq:dispbulk}
\end{equation}
whose inversion gives us the dispersion law of the \textit{bulk}
polaritons. Here $\varepsilon_{zz}$, $\mu_{xx}$, $\mu_{yy}$, and
$\mu_v$ are all functions of frequency $\omega$ (i.e. they are
functions of $k_0=\omega/c$), and $\varepsilon_{v} =
\varepsilon_{xx}+ \varepsilon_{xy}^2/\varepsilon_{yy}$ and $\mu_{v}
= \mu_{xx}+\mu_{xy}^2/\mu_{yy}$ are introduced as the Voigt relative
permittivity and relative permeability, respectively.  From
(\ref{eq:dispbulk}) also one can conclude that the asymptotic limits
of the dispersion curves for the bulk polaritons appear to be for
frequencies where $\mu_{xx} \to 0$.

In order to find the dispersion law of the \textit{surface}
polaritons, corresponding boundary conditions must be imposed to
match the components of the field vectors on both sides of the
interface between vacuum and the composite structure (we distinguish
these media with the index $i=1,2$). According to
Fig.~\ref{fig:fig_SPP} the $y$-axis is directed normal to the
interface between the media, therefore, the wavevector components,
which are responsible for the wave attenuation in both positive and
negative directions along the $y$-axis, are defined by quantities
$\kappa_i$.

Involving a pair of the divergent Maxwell's equations and the
corresponding boundary conditions the dispersion equation for the
surface polaritons at the interface between vacuum and bigyrotropic
semi-infinite media is obtained as
\begin{equation}
\kappa_1\mu_v+\kappa_2 \mu_1 + i k_x
\mu_1\frac{\mu_{xy}}{\mu_{yy}}=0. \label{eq:dispsurface}
\end{equation}
Taking into consideration the relations with respect to the
attenuation coefficients $\kappa_1$ and $\kappa_2$
 \begin{equation}
k_x^2-\kappa_1^2-k_0^2\varepsilon_1\mu_1=0. \label{eq:kappa1}
\end{equation}
\begin{equation}
\begin{split}
k_x^2&\left(\varepsilon_{xx}\mu_{xx}-\varepsilon_{xy}\mu_{xy}\right)-
\kappa_2^2\left(\varepsilon_{yy}\mu_{yy}-\varepsilon_{xy}\mu_{xy}\right)-
k_0^2\varepsilon_{yy}\mu_{yy}\varepsilon_{v}\mu_{v}\\&+ i
k_x\kappa_2\left[\mu_{xy}(\varepsilon_{yy}-\varepsilon_{xx})-
\varepsilon_{xy}(\mu_{yy}-\mu_{xx})\right] =0,
\end{split}
\label{eq:kappa2}
\end{equation}
the dispersion equation (\ref{eq:dispsurface}) can be expanded into
the biquadratic equation with respect to $k^2_x$
\cite{Tuz_MAGMA_2016}:
\begin{equation}
Ak_x^4 + Bk_x^2+C=0, \label{eq:biquadratic}
\end{equation}
where $A= Y^2+W^2$, $B=-k_0^2(2VY+\varepsilon_1\mu_1W^2)$,
$C=k_0^4V^2$, $V=p - s\varepsilon_1\mu_1(\mu_v/\mu_1)^2$,
$Y=u+s[(\mu_{xy}/\mu_{yy})^2 -(\mu_v/\mu_1)^2]-q
(\mu_{xy}/\mu_{yy})$, $W=(\mu_v/\mu_1)[2s(\mu_{xy}/\mu_{yy})-q]$,
$p=\varepsilon_{yy}\mu_{yy}\varepsilon_v\mu_v$,
$q=\varepsilon_{xy}(\mu_{yy}-\mu_{xx})-\mu_{xy}
(\varepsilon_{yy}-\varepsilon_{xx})$,
$s=\varepsilon_{yy}\mu_{yy}-\varepsilon_{xy}\mu_{xy}$,
$u=\varepsilon_{xx}\mu_{xx}-\varepsilon_{xy}\mu_{xy}$, whose
solution is trivial.

From four roots of (\ref{eq:biquadratic}) those must be selected
which satisfy the physical conditions, namely, wave attenuation as
it propagates, that imposes restrictions on the values of
$\kappa_i$, whose real parts must be positive quantities.
Furthermore, it is evident that the asymptotic limits of the
dispersion curves for the surface polaritons appear to be for
frequencies where $A \to 0$.

\begin{figure}
\centering \fbox{\includegraphics[width=6.3cm]{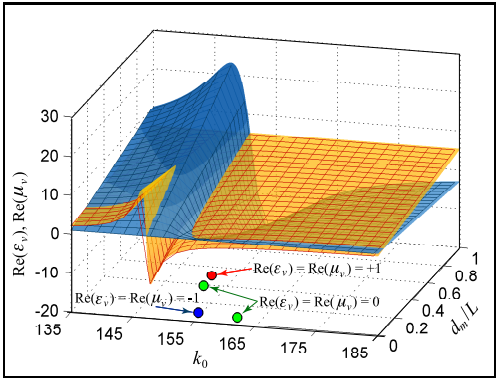}}
\caption{(Color online) Two surfaces depict behaviors of the real
parts of the Voigt relative permittivity (yellow surface) and
relative permeability (blue surface) versus wavenumber in free space
and the ratio of the layers' thicknesses. The green, red and blue
circles plotted at the figure bottom are projections of the
conditions (i)~$\epsilon_v=\mu_v=0$, (ii)~$\epsilon_v=\mu_v=+1$, and
(iii)~$\epsilon_v=\mu_v=-1$, respectively.} \label{fig:fig_NS}
\end{figure}

From the form of dispersion equations for bulk and surface
polaritons it is obvious that their spectral features substantially
depend on the dispersion characteristics of components of both
effective permeability tensor and effective permittivity tensor of
the resulting finely-stratified structure. Moreover, since the
underlying components of these tensors are all functions of the
frequency, external magnetic field strength, layers thicknesses, and
physical properties of the materials forming the superlattice, the
regions of polaritons existence are determined by the choice of the
values of the corresponding quantities. Therefore, in order to bring
together magnetic and semiconductor subsystems into a single
structure with desired characteristics, a multiparameter
optimization problem should be solved.

Based on the typical constitutive parameters which are inherent to
available materials (e.g. In$_{2-x}$Cr$_x$O$_3$, Cd$_{1-x}$Mn$_x$Te,
In$_{2-x}$Cr$_x$O$_3$, FeF$_2$/TlBr) \cite{Kussow_PhysRevB_2008,
Ta_PhotNanFundApp_2012, Aoud_IEEETera_2013}, the characteristic
resonant frequencies of magnetic and semiconductor subsystems as
well as the resulting structure period are chosen and fixed, and
then the search for an optimal fractions balance $\delta_m$ versus
$\delta_s$ ($\delta_m=d_m/L$, $\delta_s=d_s/L$,
$\delta_m+\delta_s=1$) is proceeded via altering the layers'
thicknesses within the period. As the objective function of the
optimization problem the real parts of the Voigt relative
permittivity $\varepsilon_v$ and relative permeability $\mu_v$ are
selected since they completely characterize the properties of waves
propagation through a gyroelectromagnetic medium. Three particular
cases are of interest: (i) $\varepsilon_v \to 0$; $\mu_v \to 0$;
(ii) $\varepsilon_v \to 1$; $\mu_v \to 1$; (iii)  $\varepsilon_v \to
-1$; $\mu_v \to -1$, as far as these conditions define the
asymptotic limits of the polariton branches. The graphical solution
of the discussed optimization problem is depicted in
Fig.~\ref{fig:fig_NS}, where the resolved configurations are
distinguished by the color circles. Remarkably, in all cases
$L/\lambda \approx 3 \times 10^2$.

\begin{figure}
\centering \fbox{\includegraphics[width=6.6cm]{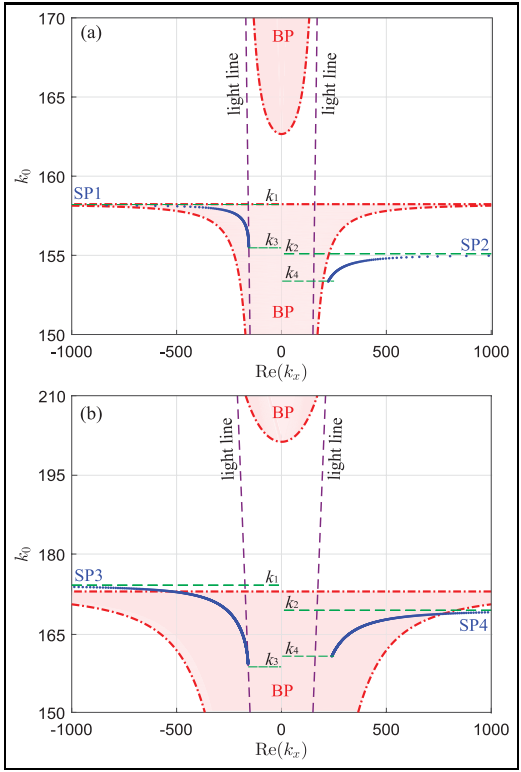}}
\caption{(Color online) Dispersion characteristics of the bulk
polaritons (red dash-dot lines) and the surface polaritons (thick
blue solid lines)  that are plotted for two chosen superlattice
configurations differed by geometric factors (a) $\delta_m = 0.081$,
$\delta_s= 0.919$, and (b) $\delta_m= 0.396$, $\delta_s = 0.604$,
which are reached under conditions $\varepsilon_v = \mu_v = 0$.}
\label{fig:fig_DC}
\end{figure}

\begin{figure}
\centering \fbox{\includegraphics[width=6.5cm]{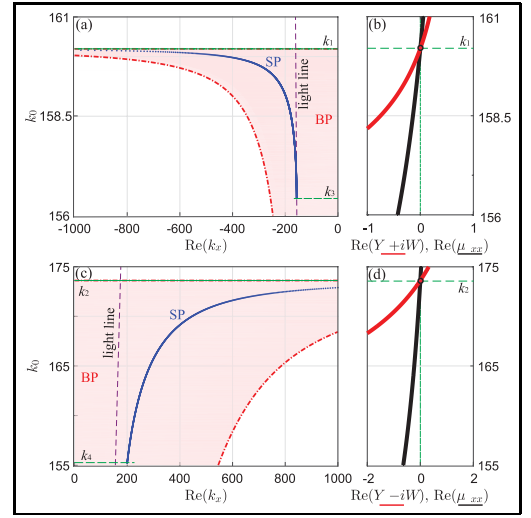}}
\caption{(Color online) (a,c) Dispersion characteristics of the bulk
(red dashed lines) and surface (blue thick lines) polaritons. (b,d)
Frequency dependencies of $Y \pm iW$ (red lines) and $\mu_{xx}$
(black lines) for two chosen superlattice configurations differed by
geometric factors (a,b) $\delta_m = 0.107$, $\delta_s = 0.893$ and
(c,d) $\delta_m = 0.551$, $\delta_s = 0.449$, which are reached
under conditions $\varepsilon_v = \mu_v = -1$ and $\varepsilon_v =
\mu_v = +1$, respectively.} \label{fig:fig_DC1}
\end{figure}

Further we consider the dispersion features of both bulk and surface
polaritons which are inherent to the composite structure
corresponding to three above specified configurations. In order to
find the real solutions related to normal modes, absence of losses
in constitutive parameters of the underlaying materials is supposed.
Solutions obtained from (\ref{eq:dispbulk}) are depicted in
Figs.~\ref{fig:fig_DC}-\ref{fig:fig_DC1} with the red dash-dot
lines, while the regions of existence of the bulk polaritons are
presented by the red colored areas denoted with the abbreviation
`BP'. There are two separated frequency regions which represent the
bulk excitations. One can see that the upper area is bounded by the
light lines and its lower limit is restricted by the line at which
$\mu_v = 0$. At the same time, the top limit of the bottom area is
at the line where $\mu_{xx} = 0$.

Still assuming that there are no losses in the constitutive
materials, four solutions of the dispersion equation
(\ref{eq:biquadratic}) with respect to the propagation constant
$k_x$ of the surface polaritons are calculated. Importantly, the
spectral characteristics of the surface polaritons in the structure
being under an action of an external static magnetic field  in the
Voigt geometry possess a nonreciprocal nature, i.e. $k_0(k_x) \ne
k_0(-k_x)$. Therefore, the appropriate root branches between these
four solutions should be properly selected in order to ensure that
they are physically correct (i.e. they provide the wave
attenuation).

Therefore, the restriction that both attenuation functions
$\kappa_i$ simultaneously are real positive quantities is imposed,
and the physically correct root branches are plotted with thick blue
solid lines in Figs.~\ref{fig:fig_DC}-\ref{fig:fig_DC1}. Here the
number next to the abbreviation `SP' denotes the particular surface
polariton branch. Thus, there is one root branch in each range of
positive $k_x$ (SP2/SP4) and negative $k_x$ (SP1/SP3) for the
corresponding investigated case.

From equation (\ref{eq:solutionbi}) it is evident that the
asymptotic limits of all these root branches are defined by the
condition $A = 0$ (i.e. $Y^2 + W^2 = 0$, implying for the lossless
system $Y$ is a real number, while $W$ is an imaginary number,
therefore, $A$ is always a real number, too). As a result the
surface polariton branches are restricted by two asymptotic
conditions: $Y - iW=0$ and $Y + iW=0$ for branches SP2/SP4 and
SP1/SP3, respectively. At these particular conditions
$\text{Re}(\kappa_i)\to \infty$ resulting in that the corresponding
penetration depths $\tau_i=1/\text{Re}(\kappa_i)$ become to be zero.
Also it should be noted that the branches SP1/SP3 are restricted by
the light line $k_0=-\omega/c$ while their lower limit is restricted
by the line at which $\text{Re} (\kappa_1)=0$ at $\omega_3=k_3c$ and
$\omega_4=k_4c$, respectively. At the same time, the lower limit of
the branches SP2/SP4 are restricted by the line at which
$\text{Re}(\kappa_2)=0$. At these lower limits the penetration depth
of the surface polariton into the appropriate medium becomes to be
infinite.

Fig.~\ref{fig:fig_DC}~(a,b) depicts the dependence of the polariton
dispersion characteristics on the choice of the structural
parameters of the superlattice for both positive and negative $k_x$
directions. Firstly, we consider the superlattice configurations
which appear at the reached condition $\varepsilon_v = \mu_v =0$.
The first found filling factors balance is: $\delta_m = 0.081$ and
$\delta_s = 0.919$, which is presented in Fig.~\ref{fig:fig_DC}~(a).
It is clear that the surface polariton branches possess
nonreciprocal nature and exist in different spectral ranges. In the
direction of negative $k_x$ the branch SP1 of the surface polariton
appears to be in the same range where the bulk polariton  exists.
Besides, in the direction of positive $k_x$ the regions of existence
of the bulk and surface polaritons (branch SP2) do not overlap at
all. Note, the branches SP1 and SP2 have different cut-off
frequencies. Fig.~\ref{fig:fig_DC}~(b) presents the dispersion
curves of the bulk and surface modes for the second found filling
factors balance: $\delta_m = 0.396$ and $\delta_s = 0.604$. As in
the previous case, the spectrum of the surface polaritons is
nonreciprocal, but the frequency ranges of branches of the surface
polaritons in positive and negative $k_x$ directions coincide.
Moreover, the branches SP3 and SP4 reemerge from the bulk continuum
at some finite values of $k_x$ and tend to the asymptotic limits
discussed above.

The regions of existence of the surface polaritons drastically
change when the superlattice configurations correspond to the
reached conditions  $\varepsilon_v = \mu_v =\pm 1$, as it is
depicted in Fig.~\ref{fig:fig_DC1}. The obtained results show that
for negative quantities of $\varepsilon_v$, $\mu_v$ the complete
coincidence of the regions of existence of the bulk polaritons and
the branches of the surface polaritons is found to be in the
negative $k_x$ direction (Fig.~\ref{fig:fig_DC1}~(a)). Besides, for
positive quantities of $\varepsilon_v$, $\mu_v$ the complete
coincidence of the mentioned characteristics appear to be in the
positive $k_x$ direction (Fig.~\ref{fig:fig_DC1}~(c)). Moreover,
there are two extreme cases when asymptotic limits of lower
BP-branch and SP-branch coincide. These conditions are $Y + iW =
\mu_{xx} = 0$ in the negative $k_x$ direction and $Y - iW = \mu_{xx}
= 0$ in the positive $k_x$ direction.

To conclude, we have examined dispersion characteristics of the bulk
and surface polaritons in the magnetic-semiconductor superlattice
which is under an action of an external static magnetic field in the
Voigt geometry. It is observed that in the case when specific
conditions related to the superlattice's constitutive parameters and
filling factor are satisfied, the regions of existence of the bulk
and surface polaritons overlap, moreover, they can have the same
asymptotic limits. We believe that such peculiarities can give great
advantages when providing excitation of the surface polaritons via
nonlinear coupling.

\bibliography{Tuz_Fesenko_OL}

\end{document}